# On the graph coloring check-digit scheme with applications to verifiable secret sharing

Kamil Kulesza, Zbigniew Kotulski
{kkulesza,zkotulsk}@ippt.gov.pl
Institute of Fundamental Technological Research
Polish Academy of Sciences
ul. Świętokrzyska 21, 00-049 Warsaw, Poland

**Keywords**: verifiable secret sharing, graph coloring, cryptography, coding theory

**Abstract.**

In the paper we apply graph vertex coloring for verification of secret shares. We start from showing how to convert any graph into the number and vice versa. Next, theoretical result concerning properties of $n$-colorable graphs is stated and proven. From this result we derive graph coloring check-digit scheme. Feasibility of proposed scheme increases with the size of the number, which digits are checked and overall probability of errors. The check-digit scheme is used to build shares verification method that does not require cooperation of the third party. It allows implementing verification structure different from the access structure. It does not depend on particular secret sharing method. It can be used as long as the secret shares can be represented by numbers or graphs.

## 1. Introduction

Graphs find applications in every field of computer science. Graph theory provides many NP class problems, hence, it is not surprising that they find applications in the cryptography and data security. Minimal vertex coloring for an arbitrary graph, is known to be NP (see [11] [13]). A very good example of public-key cryptosystem, "Polly Cracker", that makes use graph $n$-coloring, can be found in [10]. In the field of secret sharing, graphs were applied while studying access structures, for instance by Blundo, DeSantis, Stinson and Vaccaro [3].

Secret sharing allows splitting a secret into different pieces, called shares, which are given to the participants, such that only certain group (authorized set of participants) can recover the secret. Participants not forming such set should have no information about the secret. Secret sharing schemes were independently invented by George Blakley [2] and Adi Shamir [15]. Many schemes were presented since, for instance, Asmuth and Bloom [1], Brickell [4], Karin-Greene-Hellman (KGH method) [9]. In our paper we use last method in order to illustrate proposed method for shares verification.

The secret in KGH method is a vector of $\eta$ numbers $S_\eta = \{s_1, s_2, ..., s_\eta\}$. Some modulus $k$ is chosen, such that $k > \max(s_1, s_2, ..., s_\eta)$. All $t$ participants are given shares that are $\eta$-dimensional vectors $S_\eta^{(j)}$, $j = 1, 2, ..., t$ with elements in $Z_k$. To retrieve the secret they have to add the vectors component-wise in $Z_k$. For $k = 2$, KGH method works like $\oplus$ (xor) on $\eta$-bits numbers, much in the same way like Vernam one-time pad. If $t$ participants are needed to recover the secret, adding $t-1$ (or less) shares reveals no information about secret itself. Interesting feature of KGH is that when certain vectors $S_\eta^*$ are excluded (not allowed) from the set of possible secret values, method remains equally secure. Again, having $t-1$ parts (or less) of the secret reveals no information about the secret itself. KGH with excluded vectors is referred as KGHe. Certainly, for same $\eta$ (vector length) the size of the "secret space" is smaller for KGHe than for KGH.

In practice, it is often needed that only certain specified subsets of the participants should be able to recover the secret. The authorized set is a subset of secret participants that are able to recover secret. The access structure describes all the authorized subsets. To design the access structure with required capabilities, the cumulative array construction can be used, for details see, for example, [8]. Combining cumulative arrays with KGH method, one obtains implementation of general secret sharing scheme (see, e.g., [14]). While designing such an implementation, one can introduce required capabilities not only in terms of access structure but also others, like security (e.g., perfectness), see [12], [17].

The simple secret sharing schemes are not secure against cheating (e.g., some participants alter their shares). Cheating can result not only in problems with recovering the secret, but also can compromise it. For the Shamir scheme, it was shown (see [18]) that in some instances a cheating participant might submit the false secret share, that provides no information about the secret, and recover secret once all participants from an authorized subset pool their shares. The problem can be addressed by verifiable secret sharing (VSS) schemes. Then, an adequate verification algorithm allows the honest participants to detect cheating (at least, with known probability) and avoid compromising the secret. Cheating is not limited to secret participants only, but such a topic is beyond scope of this paper. Verification set of shares (VSoS) is the set of shares that are required for verification procedure to take place. Verification structure is the superset containing all verification sets of shares. The verification of the secret shares can take place in public; moreover, often the set of participants can verify the validity of their shares together. This is the case of publicly verifiable secret sharing (PVSS), for instance see [16].

Major idea of this paper to use graph's integral property such as vertex coloring for verification of secret shares. First we present method for binary strings verification and next apply it to the secret shares. Every number can be assigned corresponding graph structure and vice versa (sections 2.1 – 2.3). This allows to treat any number as the graph and test for properties associated with the graphs. Many of the graphs' properties are related to NP problems. We focus on the graph vertex $n$-coloring and discuss known results that are needed further in the paper (section 2.4). In the next section, we find and prove upper bound for maximum number of graphs, that have fixed number of vertices and can be colored with $n$ colors. Graph coloring based check-digit scheme follows in section 3. We describe and justify scheme capabilities. Check digit scheme is preliminary step to develop shares verification method. Before going that far, we devote section 4 to discuss secret sharing for the graph having in mind equivalence between graphs and numbers.

At this point all preliminaries are provided, hence we can introduce shares verification method (section 5.1). Check digit scheme serves as the engine for the method and allows to describe its security in section 5.2. Finally, we propose generalization of verification method to arbitrary numbers (section 6).

## 2. Into the realm of graph coloring

At the beginning simple scheme that allows conversion between graphs and numbers is presented. While more sophisticated methods can be used (e.g. [11]), the one chosen well illustrates development of main results.

Notation:

$G(V,E)$ is the graph, where $V$ is a set of vertices and $E$ is a set of edges, with $|E|$ edges and $|V|$ vertices; $v_i$ denotes $i$th vertex of the graph, $v_i \in V$; $K_n$ stands for the complete graph on $n$ vertices, $\chi(G)$ is chromatic number for the graph $G$.

### 2.1 Graph description

Graph $G$ is described by the square adjacency matrix $\mathbf{A} = [a_{ij}], i, j = 1, 2, ..., m$. The elements of $\mathbf{A}$ satisfy:

- for $i \neq j$, $a_{ij} = 1$ if $v_i v_j \in E$ (vertices $v_i$, $v_j$ are connected by an edge) and $a_{ij} = 0$, otherwise;

- for $i = j$, $a_{ii} = \alpha$, where $\alpha \in Z_k$ is the number of color assigned to $v_i$. In $Z_k$, $k \geq \chi(G)$ denotes the number of colors that can be used to color vertices of $G$ (in other words, $k$ is the size of the color palette).

In the case that the graph coloring is not considered, $k=1$, and all entries on **A**'s main diagonal are zero.

*Example 1*

Take the graph $G$ with 4 vertices, colored with 3 colors:

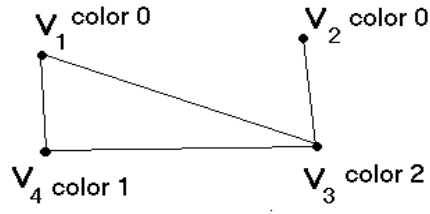

The adjacency matrix of the graph $G$ (only the graph structure, no colors) is presented on the left hand side below, while the whole adjacency matrix **A** with encoded coloring is given on the right hand side.

|       | $v_1$ | $v_2$ | $v_3$ | $v_4$ |       | $v_1$ | $v_2$ | $v_3$ | $v_4$ |
|-------|-------|-------|-------|-------|-------|-------|-------|-------|-------|
| $v_1$ | 0     | 0     | 1     | 1     | $v_1$ | 0     | 0     | 1     | 1     |
| $v_2$ | 0     | 0     | 1     | 0     | $v_2$ | 0     | 0     | 1     | 0     |
| $v_3$ | 1     | 1     | 0     | 1     | $v_3$ | 1     | 1     | 2     | 1     |
| $v_4$ | 1     | 0     | 1     | 0     | $v_4$ | 1     | 0     | 1     | 1     |

∎

Coloring and the chromatic number are integral properties of any graph. Given the graph $G$, it is always possible to find its chromatic number and $n$-coloring.

## 2.2 Coding the matrix A

**A** is a symmetric matrix, hence having all the entries on the main diagonal and all the entries below main diagonal, one can describe whole matrix (and as the result graph $G$). Thus, it can be written as the sequence $a_{21}a_{31}a_{32}a_{41}a_{42}a_{43}\ldots a_{m(m-1)}a_{11}a_{22}\ldots a_{mm}$, where the first binary part ($a_{21}a_{31}a_{32}a_{41}a_{42}a_{43}\ldots a_{m(m-1)}$) corresponds to all the entries below main diagonal, while second decimal one ($a_{11}a_{22}\ldots a_{mm}$) to the main diagonal itself.

*Example 1* (continuation)

Coding matrix **A** we obtain

| $a_{21}$ | $a_{31}$ | $a_{32}$ | $a_{41}$ | $a_{42}$ | $a_{43}$ | $a_{11}$ | $a_{22}$ | $a_{33}$ | $a_{44}$ |
|----------|----------|----------|----------|----------|----------|----------|----------|----------|----------|
| 0        | 1        | 1        | 1        | 1        | 1        | 0        | 0        | 2        | 1        |

∎

## 2.3 Coding the number as the graph

This can be done by converting number into binary form. Then the adjacency matrix **A** is encoded in the same way as depicted in the section 2.2. If the length $l$ of binary number does not yield integer solution to the equation $l = \dfrac{(m-1)m}{2}$, then lacking matrix entries can be added. Method for entries addition is further discussed in the section 3.

## 2.4 On finding graph's chromatic number and coloring

In the general case, finding graph minimal coloring or even decision on the chromatic number are of NP class (eg. see [11], [13]). Yet, it is interesting to note that usually graph can be properly colored in many ways. Problem is complex and difficult to comprehend. For instance: there are numerous examples that single edge change in the graph structure can result in dramatic change in its coloring. Any significant result in this field would approach us to resolving famous P versus NP question. Current research concentrates on three fields:
1. Algorithms. Building better approximation solution. Algorithms have polynomial time characteristics and allow to color graph with $\chi(G) = n$ with the smallest $k$-color palette ($|V| \geq k \geq n$). Results available allow to obtain quite reasonable $k$ values (e.g. see [7]). It is important to note that mentioned above algorithms can work in the deterministic way.
2. Probabilistic methods. Present state of the art is given in [13].
3. Theory. Finding new coloring criterions and new cases, that are solvable in polynomial time (e.g. see [5]). To give reader a flavor we provide few well known results, that are useful later in the paper.

**Theorem**. *Every graph G of $K_n$ configuration has $\chi(G) = n$.*

**Lemma**. *Every graph G having a subgraph of $K_n$ configuration has $\chi(G) \geq n$.*

**The Brook's theorem** (1941). *If the graph G is not an odd circuit or a complete graph, then $\chi(G) \leq d$, where d is the maximum degree of a vertex of G.*

In further considerations, two observations are used:
1. Vertex coloring of the graph can be checked only when the structure of the graph is known.
2. Checking, whether a graph is properly colored, can be done in linear time, see [11].

## 2.5 Upper bound on number of graphs with fixed $|V|$ and $n$

Let's denote $\Gamma(|V|)$ as the number of all possible graphs having $|V|$ vertices. One can calculate that:

$$\Gamma(|V|) = 2^{\frac{|V|(|V|-1)}{2}}. \qquad (1)$$

In other words, $\Gamma(|V|)$ represents number of binary sequences such that every bit corresponds to the possible edge in the $|V|$ vertices graph (1 denotes presence of the edge, 0 otherwise).

**Number of graphs on $|V|$ vertices, with vertex $n$-coloring of the graph for fixed $n$.**

The graph coloring considered in this section is merely graph $n$-coloring ($n \geq \chi(G)$) not the minimal coloring of the graph ($n = \chi(G)$). Let's denote:
$x_j$ as the number of vertices colored by $j$th color $(0 < j \leq n)$. Vertices are partitioned into $n$ sets, such that all vertices with the same color are in one set;
$V_j$ as the set of vertices colored by $j$th color $(0 < j \leq n)$, $|V_j| = x_j$.
$P$ as particular, no-degenerate partition of $|V|$ into $n$ pieces (color sets).
$P = \{x_1, x_2, ..., x_j, ..., x_n\}$ and $\sum_{j=1}^{n} x_j = |V|$;
$\Gamma(|V|, n, P)$ as the number of the graphs having $|V|$ vertices and particular $P$ (partition of $|V|$ into $n$ pieces).
$\Gamma(|V|, n) = \max_P \{\Gamma(|V|, n, P)\}$ where $\max_P$ denotes maximum value of $\Gamma(|V|, n, P)$ over of all possible partitions $P$ for given $|V|$ and $n$.

**Theorem 1 (Upper bound on number of graphs with fixed $|V|$ and $n$)**

Let $|V| = n + y$ where $y \in N$. Then

$$\Gamma(|V|) \geq 2^y \Gamma(|V|, n) \qquad (2)$$

**Outline of proof:**
First, find that

$$\Gamma(|V|, n, P) = 2^{(x_2+x_3+...+x_n)x_1 + (x_3+x_4+...+x_n)x_2 + ... + x_n x_{n-1}} \qquad (3)$$

This allows to carry out further work on $\log_2$ (base 2 logarithms).
There are two cases to be considered:
Case 1: $|V| = n$. There is only one partition $P$ (all $x_j = 1$), substitution into (3) yields $\Gamma(|V|) = \Gamma(|V|, n)$, because $y = 0$ implies $2^y = 1$.
Case 2: $|V| > n$. First properties of $\log_2(\Gamma(|V|, n, P))$ for all possible $n$ values are examined to find general formula.
Comparing corresponding terms from $\log_2(\Gamma(|V|))$ and $\log_2(\Gamma(|V|, n, P))$ one finds that

$$\log_2(\Gamma(|V|)) \geq y + \log_2(\Gamma(|V|, n, P))$$

for any $P$. When $|V| > 2n$,

$$\log_2(\Gamma(|V|)) > y + \log_2(\Gamma(|V|, n, P))$$

for any $P$. Ultimately,

$$\Gamma(|V|) \geq 2^y (\Gamma(|V|, n))$$

∎

**Remarks:**
1. Using Theorem 1 and it's proof one can derive results stating the relations between $\Gamma(|V|)$ and $\Gamma(|V|, n)$ for more specific conditions (eg. it exist $j_0$ such that $x_{j_0} = k$, $k \geq 2$).
2. Technique used to prove Theorem 1 yields the number of the graphs that can be properly colored with $n$ colors. The result includes class of graphs $G'$ such that $\chi(G') < n$. This observation can be used to calculate the number of the graphs that have fixed chromatic number $\chi(G) = n$. Authors are busy working on this field. Obtained results are not crucial for check-digit scheme outline, hence they are omitted.

### 3. Check-digit scheme based on the vertex graph coloring

To simplify procedure's description, it is assumed that there is an efficient procedure for finding graph $G$ minimal $n$-coloring ($n = \chi(G)$). We will return to this problem at the end of the section. Let us introduce the notation:
$D$ denotes number that is stored/sent, $D'$ denotes number that is read/received
$G(D), G(D')$ denotes graphs resulting from convertion of $D, D'$ into the graph form
$col(D), col(D')$ denotes minimal $n$-coloring for $G(D), G(D')$ respectively

*PROCEDURE (checks whether $D' = D$):*

Encoding
*1. Convert $D$ to $G(D)$.*
*2. Find $col(D)$.*
*3. Store/Send $D + col(D)$.*
............{Process that can modify $D$ (eg. transmission) takes place}
Decoding

*It is assumed that* $col(D') = col(D)$[1].

1. *Read/Receive* $D'$ *and convert to* $G(D')$.
2. *Is* $col(D')$ *valid* $n$-*coloring for* $G(D')$?
    *Results: No* $\rightarrow$ *error found (terminate), Yes* $\rightarrow$ *continue*
3. *Is* $G(D')$ $(n-1)$-*colorable?*
    *Result: Yes* $\rightarrow$ *error found (terminate), No* $\rightarrow$ *positive verification for* $D' = D$ ∎

## *Analysis & Discussion*

Our check-digit scheme still has the status of "work in progress". Currently we are busy working on it's proper analysis with suitable communication model. Nevertheless we provide few observations that advocate for our scheme.

1. Graph on $|V|$ vertices can be used to encode number of $l = \frac{(|V|-1)|V|}{2}$ digits, while $col(D)$ has $|V|$ elements. Hence overhead information needed to detect errors decreases rapidly with the size of the number to be checked.

2. Proposed scheme is specially suitable for situations, when number and type of errors cannot be easily foreseen. This is major difference between graph coloring based check digit scheme and the others, that successfully detect only certain specified types of error (eg. single digit error, transpositions, etc see [6]). In the proposed scheme, detection capabilities and feasibility increase with the volume of information transferred.

3. Proposed scheme can detect all $D'$, such that $col(D)$ is different from $col(D')$. As outlined in the section 2.4 even single edge change in the graph structure (corresponding to single bit change in the number) can result in change of the graph coloring. When changes in $D$ take place, $D'$ corresponds to any graph from the set of the graphs on $|V|$ vertices (the worst case of error type, that does not allow easy classification and analysis). There are $\Gamma(|V|) = 2^{\frac{|V|(|V|-1)}{2}}$ possible graphs $G(D')$ for transmitted number $D'$.

Let $p$ be the probability that $D \neq D'$ gets undetected through check-digit scheme. Define the probability $p_n$ that $col(D)$ is also proper coloring for $G(D')$, under the condition $D \neq D'$,

$$p_n = \frac{\Gamma(|V|, n)}{\Gamma(|V|)}. \qquad (4)$$

The number of $n$-colorable graphs on $|V| = n + y$ vertices decreases very quickly as $y$ increases; from (2) we obtain that $p_n \leq 2^{-y}$.

In addition coloring using $(n-1)$ colors is checked for $D'$. This eliminates all $D \neq D'$ such $\chi(G(D')) < n$. As the result $p < p_n \leq 2^{-y}$ yields $p < 2^{-y}$. Conclusions from theorem 1 (stating upper bound for $\Gamma(|V|, n)$) can be used to derive better estimates for $p$ value.

4. In order to ensure required reliability for the check digit scheme, proper value of $k$ such that $y \geq k$, should be maintained for every $D$. This can be achieved by **A** matrix extension.

---

[1] When $col(D') \neq col(D)$, error would be detected with high probability (same reasoning as for the rest of the scheme applies). For the sake of simplicity we assume that check digits ($col(D)$) are not modified during transmission. For instance, $col(D)$ can be transmitted separately via reliable channel. Analysis of possible errors in the check digits is not complicated, but this topic needs additional studies.

There are numerous techniques that can be used for this purpose. We use example to illustrate the process and its purpose.

*Example 2*

Extension is performed by adding to **A**'s bottom additional $k$ rows. These rows have carefully chosen entries and length such, that **A** remains square matrix. Added entries correspond to extra $k$ vertices added to the graph $G$, resulting in new expanded graph $G_e$. Algorithm for choosing the entries should guarantee that although number of vertices in $G_e$ is $|V|+k$, $\chi(G_e) \leq |V|$ (eg. by using Brook's theorem – see section 2.3). Only $D$ and coloring of $G_e$ need to be transmitted. Remaining part of $G_e$ would be recovered using $D'$ (algorithm for entries addition is known to both communicating parties). Coloring would be checked for the restored $G_e$. ∎

5. Now is the time to address problem of finding minimal graph coloring. As discussed in the section 2.4, in general case this is problem of NP class. In special cases there are methods for finding it's solution in polynomial time (for instance see [11], [13]). When numbers that are to be checked result in graphs from special classes, such methods can help. Otherwise efficient approximation algorithms, mentioned in section 2.4, have to be used. Instead of looking for minimal graph coloring, approximate solution can be applied. Only consequence of such approach will be reduced value of $y$, compared to minimal coloring case. This is an optimisation problem (with: efficiency of coloring approximational algorithm, graph extension algorithm and $y$ as the parameters), that should be solved for particular implementations. We continue this discussion in the section 5.2.

## 4. Secret sharing

In this section we discuss how to share secret that is the graph with unknown structure. Our presentation is informal in style, nevertheless we hope that convincing. Major objective is to prepare ground for the graph coloring based secret shares verification method, that is described in the section 5.

Equivalency between graphs and numbers should be clear at this point of the paper. Two secrets about the graph can be shared independently: the graph structure (section 4.1) and the graph coloring (section 4.2). One should note that it is impossible to verify graph coloring without knowing its structure. On the other hand number of possible graphs for any given graph coloring grows very fast with the number of vertices in the graph (see section 2.5). Shares given to the participants can consist of graph structure part, graph coloring part or both, depending on particular implementation.

### 4.1 Sharing the graph structure

As described in the section 2.2, the structure of the graph $G$ can be written as binary number $a_{21}a_{31}a_{32}a_{41}a_{42}a_{43}...a_{m(m-1)}$ derived from matrix **A**. When any graph configuration is acceptable, then to share the graph $G$ structure is equivalent to share binary number $a_{21}a_{31}a_{32}a_{41}a_{42}a_{43}...a_{m(m-1)}$ of length $l = \frac{(m-1)m}{2}$. Taking to the account that any vector is possible (it follows from any graph configuration acceptable), all results from secret sharing over Abelian groups can be used.

In this paper we use KGH ([9]) secret sharing scheme as example. Although there are more sophisticated methods (e.g. see [4], [15]), our choice allows easy comprehension and analysis of shares verification procedure presented later in the text. Easy handling of restrictions placed on the secret space is an added value. When there are restrictions placed on the graph $G$ configurations (e.g., graph has to be connected), then to share the graph $G$ structure is equivalent to share binary number $a_{21}a_{31}a_{32}a_{41}a_{42}a_{43}...a_{m(m-1)}$ of length $l = \frac{(m-1)m}{2}$, with some (known) numbers excluded from the set of possible secrets. KGH secret sharing scheme nicely illustrates these situations. Such an approach allows to apply simple reasoning of combinatorial nature for clear understanding of secret sharing method (e.g., calculating the number of connected graphs for the sequence with given length).

### 4.2 Sharing graph coloring

As described in the section 2.2 vertex coloring of the graph $G$ can be written as the vector $a_{11}a_{22}...a_{mm}$ with entries from main diagonal of the matrix **A**. Again, taking into account that any vector is possible (when any graph configuration is acceptable and graph structure is not known), all secret sharing methods suitable to share number can be applied.

It should be emphasized that, in a general case, one can share only partitioning graph's vertices into $n$ sets (proper $n$-coloring for the graph), where $n = \chi(G)$, not a particular color-to-vertex assignment. It is due to the fact, that any secret participant can modify his share by adding component-wise constant to every digit in the number. In such a case:

a. A particular color-to-vertex assignment will be modified,
b. The partitioning graph's vertices into $n$ sets (proper $n$-coloring for the graph) will remain valid,
c. Shares verification method proposed in the next section will not detect share modification.

The special case, when coloring with particular color-to-vertex assignment is securely shared, although possible, is beyond the scope of this paper.

### 5. VSS using graph coloring

Property that two pieces of information can be stored/shared in the graph can be used towards building VSS (Verifiable Secret Sharing) scheme. Secret that is shared can consist of the graph structure and coloring, or the structure only. When it comes to graph coloring, only partitioning graph's vertices into $n$ sets (partition mask) is shared. Proposed VSS is not able to detect modifications in particular color-to-vertex assignment.

### 5.1 Verification of the graph shares

The two pieces of information (the graph structure, the vertex coloring) can be shared separately. Again analysis, whether to share two pieces of information together or not, is beyond the scope of the article. For the sake of simplicity it is assumed that only graph structure is being shared, every share consist of both parts and has the form of the sequence of numbers: $a_{21}a_{31}a_{32}a_{41}a_{42}a_{43}...a_{m(m-1)}a_{11}a_{22}...a_{mm}$. Each secret participants is given such number. It is essential that secret share's graph coloring part ($a_{11}a_{22}...a_{mm}$) is not related to the structure part $a_{21}a_{31}a_{32}a_{41}a_{42}a_{43}...a_{m(m-1)}$.

An algorithm of detecting cheating participants, without co-operation of the trusted third party, will be presented. Due to the fact, that verification requires exchange of information about shares, secure communication channel (e.g., see [12]) is needed. During this discussion, it is assumed that safe protocol for information exchange is in place [2].

Definitions:

*Verification set of shares (VSoS)* is the set of shares that are required for verification procedure to take place.

*Verification structure* is the superset containing all verification sets of shares.

---

[2] In this paper the trusted third party is understood as additional trusted entity storing information (e.g. hash functions) about secret participants. Issue of secure multiparty computation (MPC) is not addressed in this paper. We assume that parties concerned use secure communication channel.

***Verification algorithm (one round).*** The participants can verify their shares without co-operation of a third party:

1. For any verification set of shares, the shares are combined to form $G(D')$ with some $n$ coloring. Structural parts of the shares are combined separately from coloring part.
2. Decoding part of check-digit procedure is applied (see section 3). It checks validity of $col(D')$ and $\chi(G(D'))$. Result of the procedure yields result of verification algorithm. ∎

**Discussion:** When all participants from VSoS combine structure parts of their shares, they obtain some $G(D')$. Combined coloring parts yields $col(D')$. Next, test whether $col(D')$ is valid for $G(D')$ is performed. We claim that tampered shares will not pass the test with very high probability. Performing multiple rounds of the presented algorithm is recommended in order to increase VSS security. Our research shows that VSS is more secure when minimal coloring for $G(D)$ is used. Analysis of possible attacks is similar to this of reliability parameters for check-digit scheme as presented in the section 3. Possible attacks are limited to random guessing or solving graph $n$-coloring (NP class problem).

In the verification algorithm we do not specify how to combine secret shares. This operation depends on the particular secret sharing scheme that is used. In order to verify shares, we only need to use check-digit scheme for combined shares. When VSoS is different from authorized set we suggest using KGH for shares verification. This certainly does not exclude using different algorithm for secret sharing.

*Example 3* of a simple (4,4) threshold sharing system using KGH. In the example KGH is used for secret sharing as well as for shares verification. The secret can be recovered only when all participants co-operate. Single verification set of shares consists of any two shares. Verification structure consists of all possible pair of shares taken from the set of four shares. In order to increase verification probability all possible combinations are checked. W use graph $G$, with vertices $v_1$ $v_2$ $v_3$. All operations are performed in $Z_4$. The pictures show the vertices assignment and the secret to be shared (a number denotes the vertex color).

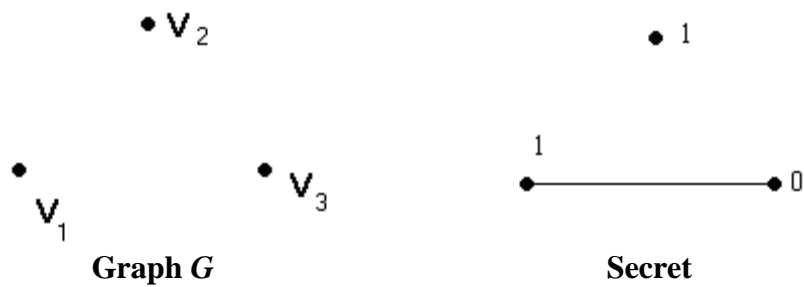

**Graph $G$**                                **Secret**

In order to illustrate the basic ideas needed for VSS, only graphs' drawings will be presented. Each drawing can be easily converted into A matrix and/or corresponding number. The four secret shares are:

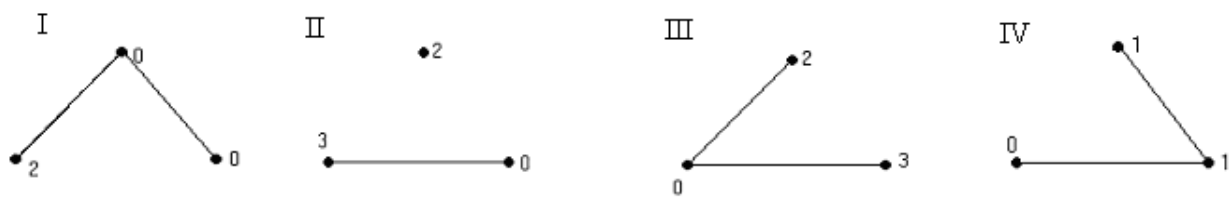

Pooling the secret shares for verification, we check all six pairs: I+II, I+III, I+IV, II+III, II+IV, III+IV.

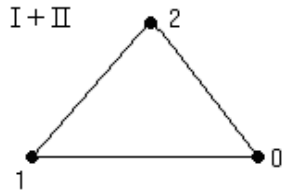
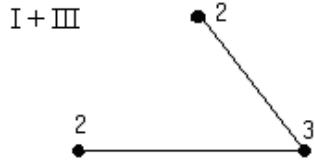
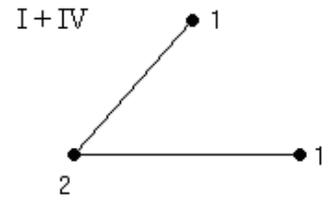
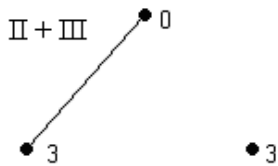
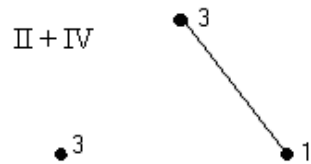
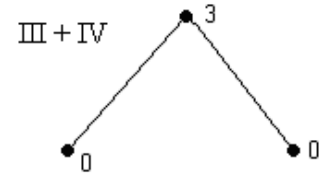

In all cases, after pooling shares, we obtained correctly colored graphs. Hence, all possible pairs of shares were checked and multiple iterations of VSS algorithm gave POSITIVE verification result in each case. Thus, the final verification result is POSITIVE. ∎

### 5.2 Discussion of verification algorithm

1. In the example 3 we used KGH to verify secret shares and to share the secret. Using same algorithm is not mandatory. For instance in example 3 one could share secret using Shamir scheme [15] and verify shares using KGH as above. In such case shares would be combined by KGH as above for verification. Once verified, shares could be combined using Shamir method to recover the secret.

2. An individual participant has information neither on the graph structure nor it's coloring. As discussed in the section 4, any graph structure is possible as the result of combining secret shares. Same reasoning applies to the graph coloring. Verification algorithm works only for participants from same VSoS. Otherwise it does not yield any information about the secret.

3. Proposed algorithm allows great flexibility in forming VSoS. Any combination of two or more shares can be used. In the VSS example given above, every two shares can be verified with each other. Other aspect of VSoS design is underlying access structure. It is important to note that any subset (with two or more elements) of authorized set can be VSoS.

4. Verificational algorithm and checking graph chromatic number.

The tools presented in the sections 2.4 and 2.5 can be used towards building graphs with required chromatic number. They also find an application to determine chromatic number for the graph. In a general case, finding chromatic number for the given graph is a NP-class problem, see [11], [13]. It can be restricted to a special cases by procedural means. Again, this problem by far exceed space limitations of this paper and deserves article on it's own. Nevertheless, two examples will be presented to illustrate a point.

### *Example 4*

Two facts are needed:
- Checking whether graph is properly colored can be done in linear time, see [11],
- Checking whether graph is 2-colourable can be done in linear time, see [11].

Assume that for the graph $G'$ formed in any round of VSS, $\chi(G') \leq 3$. When such a restriction is imposed, determination of chromatic number can be performed in linear time. ∎

*Example 5*

Assume that the graph $G'$, with $\chi(G') = n$, formed in any round of VSS, has $n$-colorable subgraph (e.g., $K_n$ type). When such a restriction is imposed, determination of chromatic number can be performed in linear time. ∎

5. A dealer that shares the secret is faced with seemingly difficult task of creating secret shares that fulfill VSS requirements. At first looks like she has to find multiple solutions to the NP class problem in order to share the secret. Much depends on particular situation (access structure and verification structure), but it seams much easier having in mind ideas presented in 2.4 and above. In addition, it is good to remind one of ($t$, $t$) threshold KGH properties, that any unauthorized set of secret participants (e.g. $t-1$ participants) can come to any solution possible, when they combine their shares. Only combination of all participants in the authorized set has to yield meaningful solution (the secret).

## 6. VSS applied to number verification

So far we discussed shares verification for graph. Having in mind graph-number equivalency, proposed secret verification method can be used otherwise: to validate any given number. Proposal will be described by mean of example:
When the secret number is shared, a graph corresponding to every individual secret share can be derived (see section 2.3). In order to use VSS protocol, the verification structure has to be designed. For instance, one can derive it from generalized access structure.
Every graph (representing single share) should have assigned colors to vertices, in order to fulfill VSS requirements. In each of the shares color part is not related to the graph structure, representing that share. Ultimately, single secret share would consist of shared number part and, possibly, shared colors part (in the absence of trusted third party). Shared colors part is a sort of checksum or hash function for the shared number. Its verification features can be activated only when verification protocol is run for some VSoS (Verification set of shares). Otherwise it remains inactive.

## 7. Concluding remarks and further research

1. Proposed shares verification method is virtually secret sharing scheme independent. It can also work with arbitrary access structure, moreover verification structure can be different from the access structure. Once general result is obtained, we are pursuing particular cases.
2. As stated in the paper feasibility of proposed verification method is based on graph $n$-coloring (problem of NP class). Although we tested issue extensively by numerous simulation, still we are busy working on formal prove of security. Because check-digit scheme underlies our verification method, hence majority of work is to be done there.
3. In this paper we presented an overview of research that is still in progress. Volume constrains forced us to simplify presentation as much as possible. As it was indicated we are busy with model of communication for our check-digit scheme. Also more sophisticated conversion methods between graphs and the numbers can be considered in order to optimize proposed check-digit scheme.
4. More methods for matrix **A** extension have to be designed and tested.
5. We have shown that once graph is coded as the number, information about it can be easily shared using results from secret sharing over Abelian groups. Once mapping between sharing information about the graph and the particular method (say KGH) is established, all previous results for the method (e.g. generalized secret sharing scheme) apply. So, in order to obtain required secret sharing implementation one does not need to work with the graphs, he simply works with numbers and uses all toolbox available for chosen method.
6. We consider using graph $n$-coloring to accommodate further extended capabilities. For instance: proposed VSS can be used as PVSS. One can design protocol that at least part of verification procedure can take place in public. Even when public communication would disclose too much, still it is possible that no secret will be maintained between participants of the round in verification protocol. For proper design of PVSS (Publicly Verifiable Secret Sharing) one should take into account constrains like VSoS-es, verification structure and number of verification protocol's rounds.